\documentclass[twocolumn,aps,reprint,showpacs,prl]{revtex4}
\usepackage{amsmath}
\usepackage{graphicx}
\usepackage{epstopdf}
\usepackage{dcolumn}
\usepackage{bm}
\usepackage{float}

\begin{document}

\title{Dynamics of a suspension of interacting yolk-shell  particles}
\author{L. E. S\'anchez D\'iaz{$^\dagger$}, E. C.
Cortes-Morales{$^\ddagger$}, X. Li{$^\dagger$}, Wei-Ren Chen{$^\dagger$}, and M. Medina-Noyola{$^\ddagger$} }

\address{$^\dagger$Biology and Soft Matter Division, Oak Ridge
National Laboratory, Oak Ridge, Tennessee 37831, USA}
\address{$^\ddagger$Instituto de F\'{\i}sica {\sl ``Manuel Sandoval
Vallarta"}, Universidad Aut\'{o}noma de San Luis Potos\'{\i},
\'{A}lvaro Obreg\'{o}n 64, 78000 San Luis Potos\'{\i}, SLP,
M\'{e}xico}
\date{\today}

\begin{abstract}

In this work we study the self-diffusion properties of a liquid of
hollow spherical particles (shells) bearing a smaller solid sphere
in their interior (yolks). We model this system using purely
repulsive hard-body interactions between all (shell and yolk)
particles, but assume the presence of a background ideal solvent
such that all the particles execute free Brownian motion between
collisions, characterized by short-time self-diffusion
coefficients $D^0_s$ for the shells and $D^0_y$ for the yolks.
Using a softened version of these interparticle potentials we
perform Brownian dynamics simulations to determine the mean
squared displacement and intermediate scattering function of the
yolk-shell complex. These results can be understood in terms of a
set of effective Langevin equations for the $N$ interacting shell
particles, pre-averaged over the  yolks' degrees of freedom, from
which an approximate self-consistent description of the simulated
self-diffusion properties can be derived. Here we compare the
theoretical and simulated results between them, and with the
results for the same system in the absence of yolks. We find that
the yolks, which have no effect on the shell-shell static
structure, influence the dynamic properties in a predictable
manner, fully captured by the theory.

\end{abstract}


\maketitle

\section{Introduction}

In recent years, there has been a growing interest in designing
and manufacturing new materials with nano and microparticles
having specifically  tailored morphologies, such as core-shell and
hollow structures \cite{lou,joo,yin}. A particularly interesting
morphology refers to yolk-shell particles, in which a hollow shell
carries a smaller particle in its interior \cite{lou,kam}.
Nanostructures of this type can be synthesized for application in
various fields such as nanoreactors \cite{kam}, lithium-ion
batteries \cite{liu}, biomedical imaging\cite{liu},
catalysis\cite{li}, and energy storage devices \cite{liu}. An
essential aspect of the technological use of yolk-shell systems
refers to the understanding of their structural and dynamic
properties in terms of the yolk-yolk, yolk-shell, and shell-shell
direct interaction forces. For example, besides the obvious steric
forces that keep a yolk inside its shell, in concentrated samples
one may have to consider the interactions between yolk-shell
complexes. A basic and elementary question, for example, refers to
the effect of the yolk-shell interaction on the Brownian motion of
the complex, as exhibited by the difference between the Brownian
motion of an empty shell, and of a shell carrying its yolk. A
second question refers to the effects of the interactions between
yolk-shell complexes on the individual and collective diffusion of
interacting yolk-shell complexes.

As a first step to address these questions, in this work we
consider a simple model representation of this class of materials,
namely, a monodisperse colloidal suspension formed by $N$ rigid
spherical shell particles of outer (inner) diameter $\sigma_s \
(\sigma_{in})$ in a volume $V$, each of which bears one smaller
(``yolk'') rigid particle of diameter $\sigma_y\ (<\sigma_{in}$)
diffusing in its interior. Here we perform Brownian dynamics
simulations to describe how the mean squared displacement (MSD) $
W(t) = <[\Delta \textbf{R}(t)]^2>/6$ and the self-intermediate
scattering function (self-ISF) $F_S(k,t)\equiv \left\langle \exp
{[i{\bf k}\cdot \Delta{\bf R}(t)]} \right\rangle $ of tagged
yolk-shell particles are influenced by the combined effect of the
interaction of the shells with their own yolks and with the other
shells, as we vary the concentration $n\equiv N/V$ of yolk-shell
particles. We then interpret the results in terms of a model
liquid of effective interacting shells, whose Brownian motion is
``renormalized'' by a time-dependent added friction that results
from averaging over the yolks degrees of freedom. This
interpretation is developed in the framework of the
self-consistent generalized Langevin equation (SCGLE) theory of
colloid dynamics, which is adapted here to the context of a
suspension of Brownian yolk-shell particles.

As our fundamental starting point, let us consider a model
monodisperse yolk-shell colloidal suspension formed by $N$
spherical shell particles in a volume $V$, each bearing one
smaller yolk particle diffusing in its interior. Let us neglect
hydrodynamic interactions and denote by ${\bf x}_{i}(t)$ and ${\bf
v}_{i}(t)$ the position and velocity of the center of mass of the
$i$th shell ($1\le i\le N$) or yolk ($N+1\le i\le 2N$) particle.
Let us denote by $M_s$ and $M_y$ the mass of, respectively, the
shell and the yolk particles, and by $\zeta^0_s$ and $\zeta^0_y$
their respective short-time friction coefficients. Then the
postulated microscopic dynamics of these $2N$ Brownian particles
is described by the following $2N$ Langevin equations,
\begin{equation}
M_i\frac{d{\bf v}_{i}(t)}{dt}= -\zeta^0_i{\bf v}_{i}(t)+{\bf
f}^{0} _{i}(t)+{\bf F}_{i}(t), \label{eq1}
\end{equation}
with  $M_i=M_s$ and $\zeta^0_i=\zeta^0_s$ for $1\le i \le N$, and
$M_i=M_y$ and $\zeta^0_i=\zeta^0_y$ for $N+1\le i \le 2N$. In
these equations, ${\bf f}^{0}_{i}(t)$ is a Gaussian white random
force of zero mean, and variance given by  $\langle {\bf
f}^0_{i}(t){\bf f}^0_{j}(0)\rangle =k_{B}T\zeta^0_i2\delta
(t)\delta _{ij}{\bf I}$ (with {\bf I} being the $3\times 3$ unit
tensor, $T$ the temperature, and $k_B$ Boltzmann's constant) and
${\bf F}_{i}(t)$ is the force exerted by all the shells and yolks
on the $i$th particle at time $t$. These forces are assumed to be
pairwise additive and determined by the radially symmetric pair
potentials $u_{ss}(r)$, $u_{sy}(r) =u_{ys}(r)$, and $u_{yy}(r)$
describing, respectively, the shell-shell, shell-yolk, and
yolk-yolk direct interactions. For concreteness, here we shall
have in mind a yolk-shell model in which the yolks only interact
with their own shells, so that $u_{yy}(r)=0$. Furthermore, the
specific case analyzed below will involve purely repulsive
hard-body interactions, defined by the conditions $\exp [-\beta
u_{ss}(r)]=H (r-\sigma_{s})$ and $\exp [-\beta u_{ys}(r)]=H
((\sigma_{in} - \sigma_y)/2-r)$, where $H(x)$ is Heaviside's step
function and $\beta\equiv 1/ k_BT$.

Our Brownian dynamics simulations consist essentially of the
numerical solution of the overdamped version of these $2N$
stochastic Langevin equations, in which one neglects the inertial
terms $M_s[d{\bf v}_{i}(t)/dt]$, according to the algorithm
proposed by Ermak and McCammon \cite{ermak}. We have used this
algorithm in the efficient, low-memory version proposed in Ref.
\cite{dub} to calculate the dynamic properties of the system
above. The results will be employed to assess the numerical
accuracy of the statistical mechanical description provided by the
self-consistent generalized Langevin equation (SCGLE) theory of
colloid dynamics \cite{scgle0, scgle1, scgle2, rmf,
todos1,todos2}, adapted here to the context of a suspension of
Brownian yolk-shell particles. This theory also starts from the
same microscopic dynamics represented by the $2N$ Langevin
equations in Eqs. (\ref{eq1}), but its general strategy  is to
first average out the degrees of freedom of the yolk particles,
which is equivalent to ``solving'' Eqs. (\ref{eq1}) for $N+1\le i
\le 2N$, i.e., for the positions and velocities of the yolk
particles, and substituting the solution in Eqs. (\ref{eq1}) for
$1\le i \le N$. Such procedure, detailed elsewhere \cite{faraday}, yields the
following set of $N$ ``renormalized'' Langevin equations involving
the positions and velocities of only the shell particles
\begin{equation}
M_s{\frac{d{\bf v}_{i}(t)}{dt}}= -\zeta^0_{s}{\bf v}_{i}(t)-
\int_0^t dt'\Delta \zeta_y(t-t')  {\bf v}_{i}(t')+{\bf f}
_{i}(t)+{\bf F} _{i}(t) \label{eq2ppp}
\end{equation}
for $i=1,2,\ldots ,N$, where now ${\bf F} _{i}(t)$ only involves
shell-shell interactions and the random force ${\bf f}_{i} (t)$
has zero mean and variance given by $ \langle {\bf f}_{i}(t){\bf
f}_{j}(0)\rangle =k_{B}T[\zeta^0_{s}2\delta (t)+ \Delta \zeta
_y(t) ]\delta _{ij}{\bf I}$. Within a reasonable set of
approximations, the time-dependent friction function $\Delta
\zeta_y (t) $ can be written as
\begin{equation}
\Delta\zeta_y(t)= \frac{k_BT n_0}{3(2\pi)^3}\int d^3 k
[kg_{ys}(k)]^2 e^{-k^2D^0_y t} F_S(k,t), \label{dzdty4s}
\end{equation}
with  $g_{ys}(k)$ being the Fourier transform of $g_{ys}(r)
\equiv\exp[-\beta u_{ys}(r)]$ and $n_0\equiv1/ \int \exp[-\beta
u_{ys}(r)] d^3r$.

The second stage in this strategy is to start from this set of $N$
renormalized Langevin equations to derive a microscopically-based
description of the macroscopic dynamic properties of the
yolk-shell Brownian liquid. We do this by adequately extending the
SCGLE theory of colloid dynamics. Omitting the details of the
derivations and approximations, the end result is a set of three
coupled approximate equations. The first of these is an approximate
expression for the time-dependent friction function $\Delta
\zeta_s (t) $ representing the friction on a tracer shell particle
due to its direct interactions with the other shells, namely,
\begin{equation} \Delta \zeta_s (t)=\frac{k_BT}{3\left( 2\pi
\right) ^{3} n}\int d {\bf k}\left[\frac{ k[S(k)-1]}{S(k)}\right]
^{2}F(k,t)F_{S}(k,t), \label{dzdt}
\end{equation}
where $n\equiv N/V$, $S(k)$ is the (shell-shell) static structure
factor, and $F(k,t)\equiv (1/N)\left\langle \sum_{i,j=1}^N e^
{{[i{\bf k}\cdot({\bf r}_i(t)-{\bf r}_j(0))]}}\right\rangle $ is
the collective intermediate scattering function. The other two
equations are approximate expressions for $F(k,t)$ and $F_S (k,
t)$, which in Laplace space read
\begin{equation}
F(k,z)=\frac{S(k)}{z+\frac{k^{2}S^{-1}(k)D^0_s }{1+\Delta
\zeta_y^* (z) +\lambda(k)\Delta \zeta_s^* (z)}}  \label{fk}
\end{equation}
and
\begin{equation}
F_S(k,z)=\frac{1}{z+\frac{k^{2}D^0_s }{1+\Delta \zeta_y^* (z)
+\lambda(k)\Delta \zeta_s^* (z)}},  \label{fks}
\end{equation}
with $\lambda (k)$ given by  \cite{todos2}
\begin{equation}
\lambda _\alpha(k)=1/[1+( k/k_{c})]^{2}, \label{lambdadkuniform}
\end{equation}
and $k_{c}= 1.305 (2\pi/\sigma)$ being an empirically-chosen
cutoff wave-vector. This specific choice results from a previous
calibration with the properties of hard-sphere systems
\cite{overdampedatomic}.

We have solved Eqs. (\ref{dzdty4s})-(\ref{lambdadkuniform}) for
our yolk-shell model above, for given  static structural
properties $g_{ys}(k)$ and $S(k)$, with the  $S(k)$ provided by
the Percus-Yevick \cite{percus} approximation with its Verlet-Weis
correction \cite{verlet}. From this solution, all the collective
and self-dynamic properties are determined, which we illustrate
here with the results for $ W(t)$ and $F_S(k,t)$. These properties
were also determined in our BD simulations and performed in a
conventional fashion \cite{allen}, except that in practice we
followed the method introduced in Refs. \cite{guev,lety}, which
involves a softened version of the potential above to describe the
interactions among yolk and shell particles, since the
BD algorithm is only rigorously defined for continuous pair potentials. The
details of these simulations, as well as the derivation of Eqs.
(\ref{dzdty4s})-(\ref{lambdadkuniform}), are provided elsewhere
\cite{paperlargo}. Let us mention that all the results discussed
in this paper correspond to a fixed yolk-shell geometry, in which
the thickness $(\sigma_s-\sigma_{in})$ of the shell is 5\%  of the
shell's outer diameter, $\sigma_{in}/\sigma_s=0.9$, and in which the yolk's
diameter is 0.2 in units of $\sigma_s$, i.e.,
$\sigma_{y}/\sigma_s=0.2$. The results reported below will be
expressed using $\sigma_s$ and $\sigma^2_s/D^0_s$ as the units of
length and time, respectively.

\begin{figure}
\includegraphics[width=0.35\textwidth]{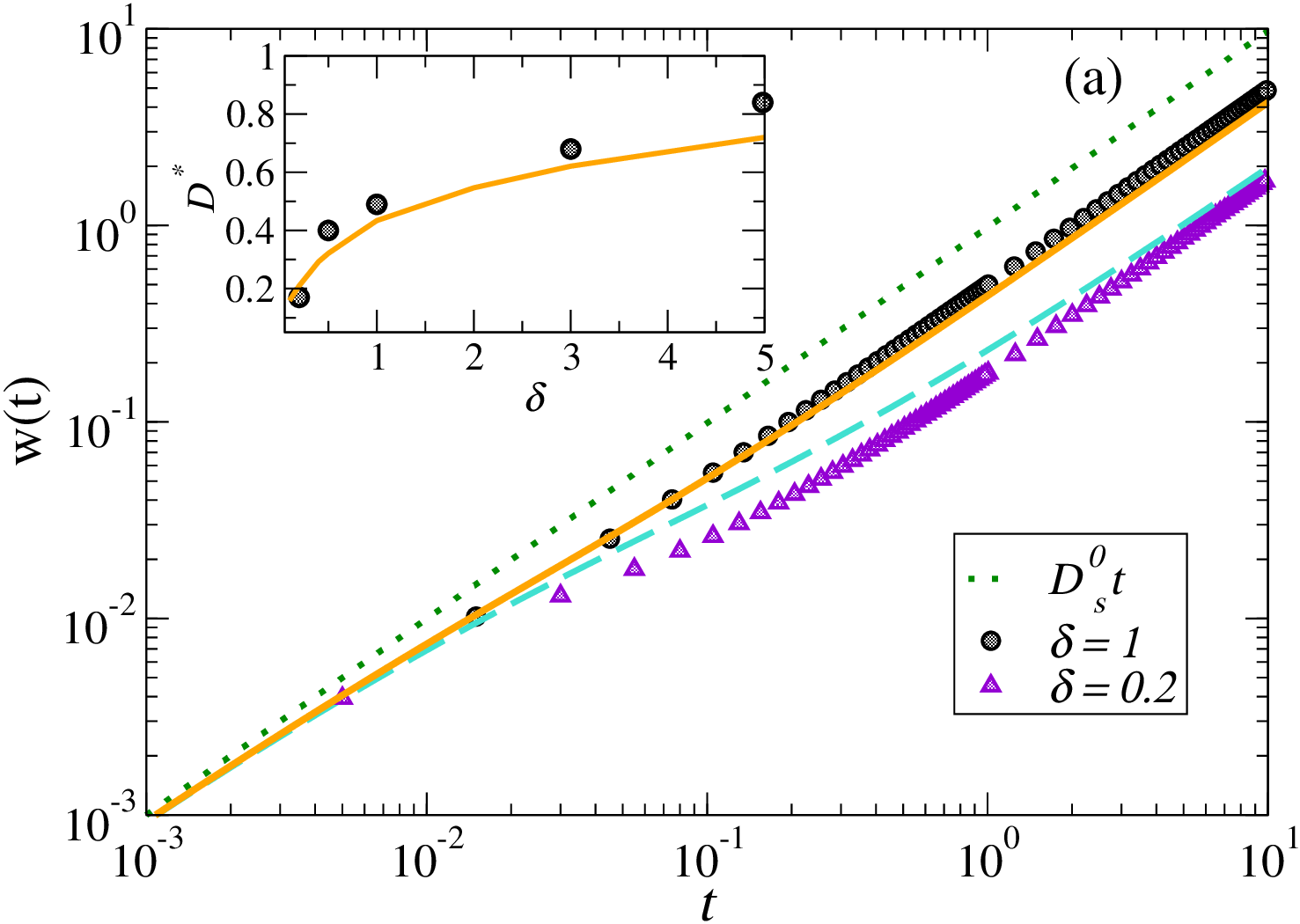}\vskip.8cm
\includegraphics[width=0.35\textwidth]{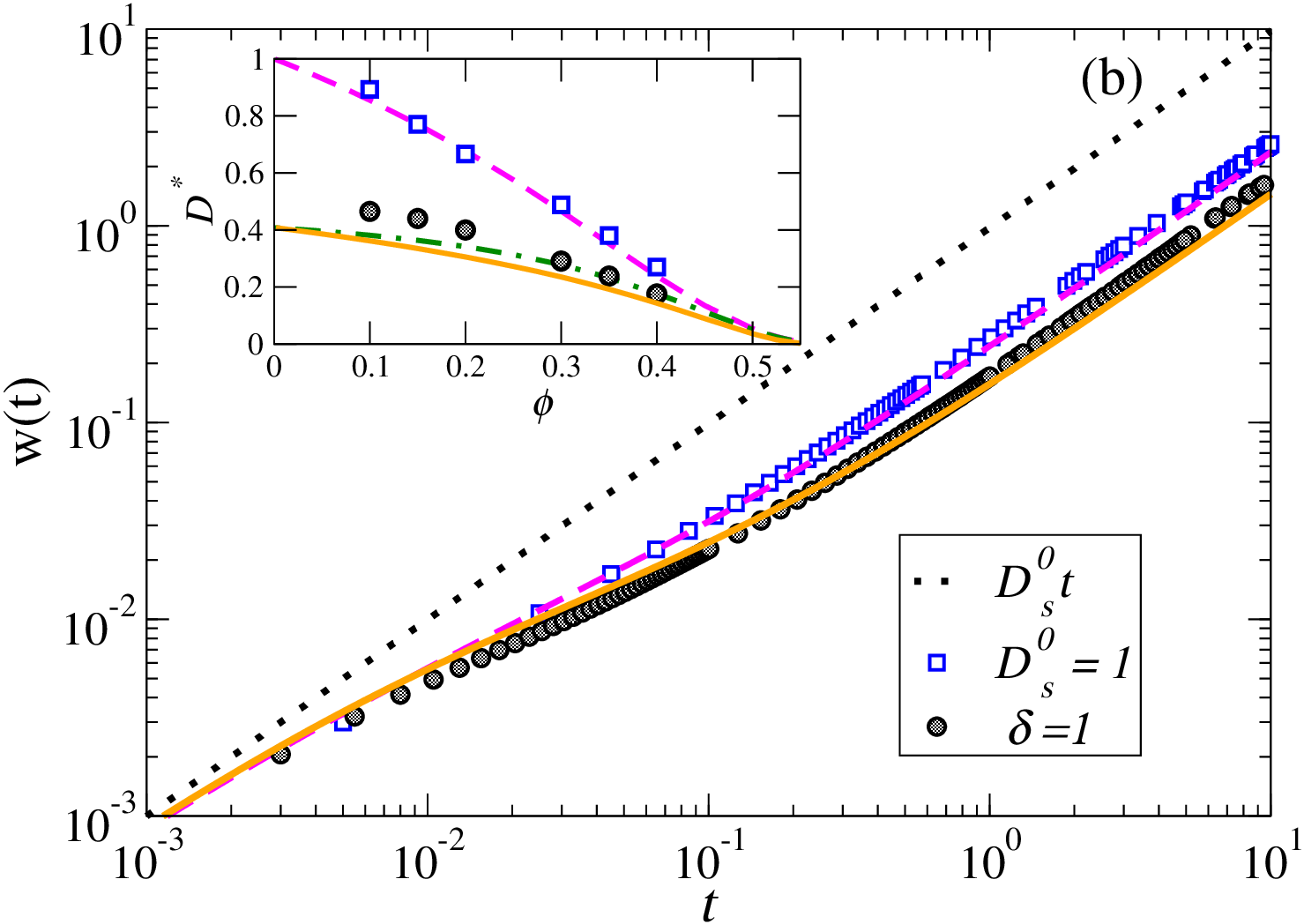}\vskip.8cm
\includegraphics[width=0.35\textwidth]{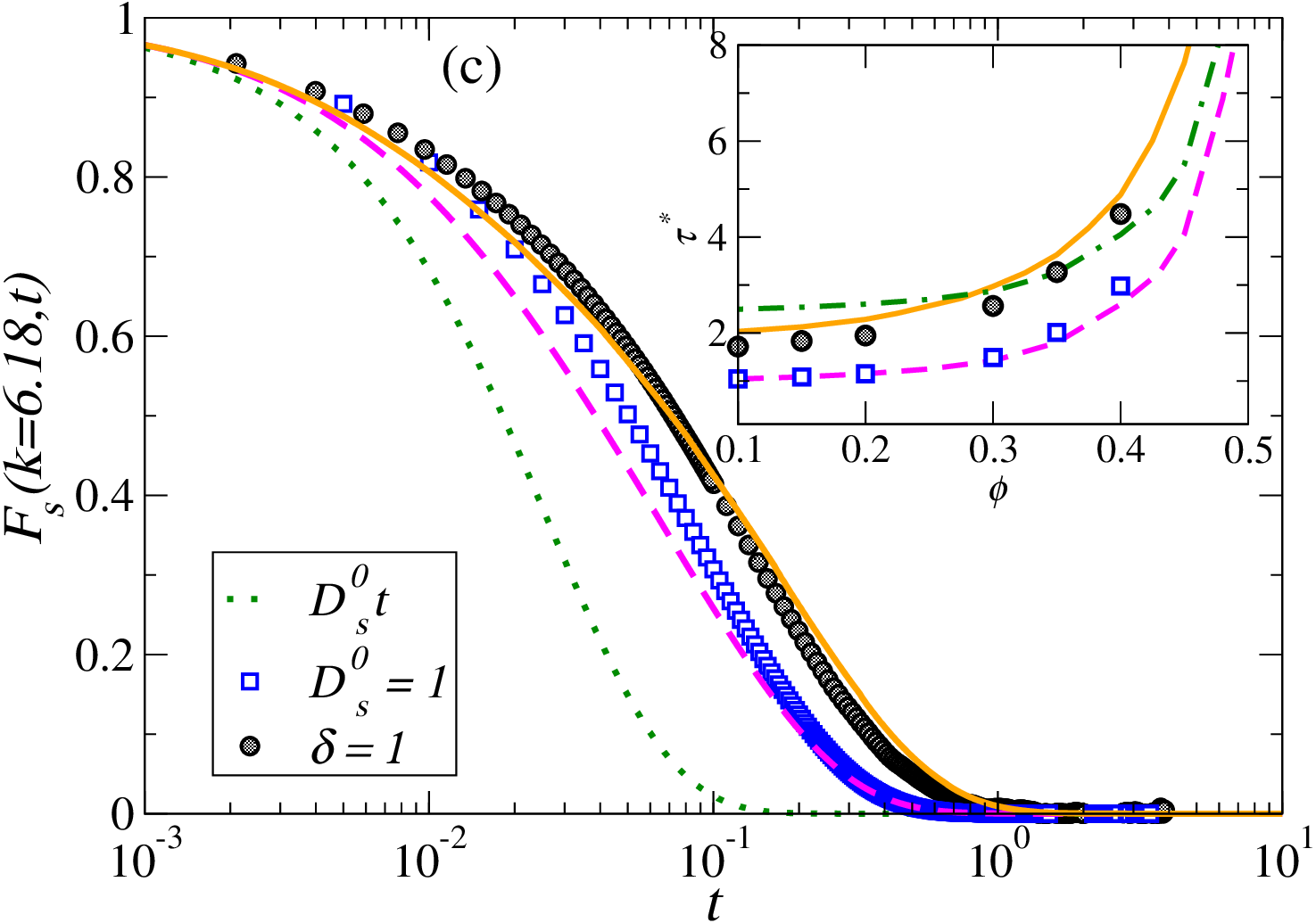}
\caption{\small  (a) Mean square displacement $W(t;\phi=0)$ of non-interacting yolk-shell particles, (b)
the MSD $W(t;\phi=0.4)$, and (c) the self-intermediate scattering function $F_s(k=6.18,t;\phi=0.4)$ 
at volume fraction $\phi=0.4$. For all figures, shell thickness 
$(\sigma_s-\sigma_{in})/2=0.05$ (or $\sigma_{in}=0.9$) and yolk
diameter $\sigma_y=0.2$. Recall that we take $\sigma_s$
as the units of length and $\sigma^2_s/D^0_s$ as the time unit.
For each figure, circles plot the Brownian
dynamics data for the yolk-shell particles, whereas the solid line is the corresponding theoretical prediction of
the SCGLE theory with a dynamic asymmetry parameter of $\delta\equiv D^0_y/D^0_s=1$. 
In figures (a) and (b) the dotted line represents the MSD $W_0(t;\phi=0)=D^0_{s}t$ of a freely diffusing empty shell,
 and in (c) it represents the self-ISF $F^s_0(k,t;\phi=0)= exp(- k^2D^0_s t)$ for the same. 
 In (a), the triangles plot the Brownian dynamics data and the dashed line plots the theoretical prediction,
 both corresponding to $\delta=0.2$. 
 In the case of (b) and (c), the squares are the simulation data for the empty shell particles, and the dashed line is 
 the corresponding theoretical prediction of the SCGLE theory with short-time diffusion coefficient $D^0_s=1$. 
The inset of (a) shows the long-time 
self-diffusion $D^*$ as function $\delta$. The insets in figures (b) and (c) are the long-time 
self-diffusion $D^*$ and $\tau^*(\equiv
k^2D_s^0\tau_\alpha)$ respectively, both as function $\phi$. Symbols and lines for these insets 
are the same as described in their own main 
figures. The dot-dashed line in inset (b) is the prediction for $D^*$ using eq. (8), and (c) shows 
the same prediction for the value of $\tau^*$. } \label{wdt1}
\end{figure}

We begin by analyzing the results in Fig. 1 of our Brownian
Dynamics simulations for the mean squared displacement
$W(t;\phi,\delta)$ of tagged yolk-shell particles, varying the
shell volume fraction $\phi\equiv \pi n \sigma_s^3/6$ and the
dynamic asymmetry parameter $\delta\equiv D^0_y/D^0_s$. These
results will be compared with the mean squared displacement
$W^{(HS)}(t;\phi)$ of the corresponding system of \emph{empty}
shells, equivalent in our model to the MSD of a Brownian
hard-sphere liquid at volume fraction $\phi$ \cite{guev,lety}.
Fig. 1(a) and 1(b) contain, respectively, the results for
$W(t;\phi,\delta)$ at $\phi=0$ (freely-diffusing yolk-shells) and
$\phi=0.4$ (strongly-interacting yolk-shells). Fig. 1(a) is intended
to illustrate two effects in a simple manner. The first is the
effect of yolk-shell interactions, which is illustrated by the comparison
of the circles, representing the MSD $W(t;\phi=0,\delta=1)$ of
freely diffusing yolk-shell particles, and  the dotted line,
representing the MSD $W_0(t)\equiv D^0_{s}t$ of freely diffusing
\emph{empty} shells. The deviation of the simulation data from
$W_0(t)$ is a measure of the additional friction effects upon the
displacement of the yolk-shell complex due to the yolk-shell
interaction. The solid line that lies near the circles represents the
prediction of our SCGLE theory. The first conclusion that we can
draw from this comparison is that the SCGLE-predicted deviation of
$W(t;\phi=0,\delta=1)$ from $W_0(t)$, coincides very
satisfactorily with the deviation observed in the BD data.

The second effect illustrated by Fig. 1(a) involves the dynamic
asymmetry parameter.  The next conclusion to draw from the
results in this figure is that the deviation of
$W(t;\phi=0,\delta)$ from $W_0(t)$ increases when the dynamic
contrast parameter $\delta$ decreases. This is illustrated by the
comparison of the BD simulations corresponding to $\delta=0.2$
(triangles) with the BD data corresponding to $\delta=1.0$
(circles). This means, for example, that if the interior of the
shell becomes more viscous, so that the ratio $\delta$ decreases,
then also the overall diffusivity of the yolk-shell particle will
decrease. As evidenced by the solid and dashed lines in Fig.
\ref{wdt1}(a), this trend is also predicted by the SCGLE theory 
and shows good quantitative agreement with the simulation data. In fact,
our theory predicts, and the simulations corroborate, that this
trend is reversed when one considers the opposite limit, in which
$\delta$ is now larger than 1. This trend is best illustrated in
the inset of Fig. \ref{wdt1}(a), which exhibits the measured
(circles) and predicted (solid line) dependence on $\delta$ of the
scaled long-time self-diffusion coefficient
$D^*(\phi,\delta)\equiv D_L/D_s^0$, obtained as
$D^*(\phi,\delta)=\lim_{t \to \infty} W(t;\phi,\delta)/D^0_st$.
There we can see that the reduction of the mobility
$D^*(\phi=0,\delta)$ from its unit value $D^*_{HS}(\phi=0)=1$, in
the absence of yolks, may be considerable. For example, for
$\delta=0.2$ we have that $D^*(\phi=0,\delta) \approx 0.17$. As a
reference, a reduction in $D^*_{HS}(\phi)$ of a similar magnitude
can also be produced as a result of pure shell-shell interactions,
but only at shell volume fractions above 40\%, as gathered from
the results of the inset in Fig. 1(b), which discusses the effects
of shell-shell interactions.

Let us now study the effects of shell-shell interactions,
suppressed in the previous discussion, by analyzing the results of
our simulations in Fig. 1(b), in which we now fix the dynamic
asymmetry parameter at the value $\delta=1$. The circles in this
figure represent the BD results for the MSD $W(t;\phi=0.4,\delta)$
of yolk-shell particles at a volume fraction $\phi=0.4$. These
results are to be compared with the squares, which correspond to
the BD results for the MSD $W^{(HS)}(t;\phi=0.4)$ of a liquid of
empty shells (or solid hard-spheres) at the same volume fraction
and same short-time self-diffusion coefficient $D^0_s$. The MSD
$W_0(t)=D^0_{s}t$ of freely-diffusing empty shells is also plotted
for reference as a dotted line. We observe that the deviation of
$W(t;\phi=0.4,\delta)$ from $W^{(HS)}(t;\phi=0.4)$ remains rather
similar to the corresponding deviation observed at $\phi=0$ in
Fig. 1(a) (where $W^{(HS)}(t;\phi=0)=D_s^0t$), but now both
$W(t;\phi=0.4,\delta)$ and $W^{(HS)}(t;\phi=0.4)$ deviate
dramatically from this free-diffusion limit. This means that for
this concentration, the mutual friction effects due to shell-shell
interactions overwhelm the ``internal'' friction effects caused by
yolk-shell interactions.

From the long-time BD data for $W(t;\phi=0.4,\delta)$ and
$W^{(HS)}(t;\phi=0.4)$ in Fig. 1(b), we can extract the value of
$D^*(\phi,\delta)$ and $D^*_{HS}(\phi)$, which represent the
mobility of a tracer yolk-shell particle and of an empty shell,
respectively. In the inset of Fig. 1(b), we plot the values of
$D^*(\phi,\delta)$ and $D^*_{HS}(\phi)$ (circles and squares,
respectively) determined from the corresponding BD results at a
few volume fractions. This inset thus summarizes the main trends
illustrated by the results in Figs 1.a and 1.b, by evidencing that
at low volume fractions the difference between the mobility of a
yolk-shell complex and the mobility of an empty shell, is
determined only by the yolk-shell friction, whereas at higher
concentrations it is dominated by the shell-shell interactions.
The solid lines in Fig. 1(b) represent again the predictions of
the SCGLE theory for the properties of the yolk-shell system,
whereas the dashed lines are the corresponding predictions for the
empty-shell (or hard-sphere) suspension. Once again, the agreement
with the simulation results is also quite reasonable for a theory
with no adjustable parameters.

Beyond this quantitative observation, however, the theoretical
description provides additional insights on the interpretation of
the qualitative trends exhibited by the simulation data of the
long-time self-diffusion coefficients $D^*(\phi,\delta)$ and
$D^*_{HS}(\phi)$. For example, it is not difficult to demonstrate
that if the shell-shell mutual friction $ \int^{\infty}_0 dt\Delta
\zeta^*_s(t;\phi,\delta)$ does not depend strongly on the presence
or absence of the yolk (which one expects to be the case at high
concentrations), then an approximate relationship between
$D^*(\phi,\delta)$ and $D^*_{HS}(\phi)$ can be derived, namely,
\begin{equation}\label{dleyhs}
D^*(\phi,\delta) = \frac{D^*_0(\delta)\times
D^*_{HS}(\phi)}{D^*_{HS}(\phi)+ D^*_0(\delta)[1-D^*_{HS}(\phi)]}.
\end{equation}
where $D^*_0(\delta)\equiv D^*(\phi=0,\delta) = [1 +
\Delta\zeta_y^*(\delta)]^{-1}$, with
$\Delta\zeta_y^*(\delta)\equiv \int^{\infty}_0 dt\Delta
\zeta^*_y(t;\phi=0,\delta)$. This expression interpolates
$D^*(\phi,\delta)$ between its exact low and high concentration
limits $D^*_0(\delta)$ and $D^*_{HS}(\phi)$, and the dot-dashed
line in the inset of Fig. 1(b) is the result of using this
approximate expression.

The BD simulations and the SCGLE theory provide other relevant
collective and self diffusion dynamic properties, such as the
intermediate scattering functions, specially amenable to
determination by dynamic light scattering techniques and adequate
index-matching methods. To close this illustrative presentation,
let us discuss the BD results in Fig. 3.c for the self-ISF
$F_S(k=6.18,t;\phi=0.4,\delta=1)$ (circles), which we compare with
the self-ISF $F_S^{HS}(k=6.18,t;\phi=0.4)$ of a liquid of
empty-shells at the same volume fraction (squares). For reference,
we also plot as a dotted line the self-ISF $F^0_S(k,t)\equiv
F_S^{HS}(k,t;\phi=0)= exp(- k^2 D^0_s t)$ of freely-diffusing
empty shells. Here too, the solid and dashed lines correspond to
the solution of Eqs. (\ref{dzdty4s})-(\ref{lambdadkuniform}) with
and without the shell friction term $\Delta \zeta_y^* (t)$, and
 comparison again indicates very
reasonable agreement with the simulation data.

In this case, the difference between the yolk-shell and empty-shell
results can also be expressed more economically in terms of the
corresponding $\alpha$-relaxation times $\tau_\alpha$, defined by
the condition $F_S(k,\tau_\alpha)=1/e$ and scaled as $\tau^*\equiv
k^2D_s^0\tau_\alpha$. The inset of Fig. 1(c) exhibits the
theoretical (solid line) and simulated (circles) results for the
yolk-shell $\tau^*(k;\phi,\delta)$ evaluated at $k=6.18$ for
$\delta=1$ as a function of $\phi$. These results may be compared
with the theoretical (dashed line) and simulated (squares) results
for $\tau^*_{HS}(k=6.18;\phi=0.4)$, corresponding to the
empty-shell suspension, with similar conclusions as in Fig. 1(b).
In analogy with the relationship in Eq. (\ref{dleyhs}), from the
SCGLE equations one can also derive an approximate relationship
between $\tau^*(k;\phi,\delta)$ and $\tau^*_{HS}(k;\phi)$, namely,
$\tau^*(k;\phi,\delta)\approx \tau^*_{HS}(k;\phi) +
\Delta\zeta_y^*(\delta)$. This prediction of the value of
$\tau^*(k;\phi,\delta)$ has a rather modest quantitative accuracy,
as indicated by the dot-dashed line in the inset. Still, it
contributes to a simple and correct qualitative understanding of
the main features of the properties of the yolk-shell system being
studied.

In summary, in this work we have carried out BD simulations and
have proposed a statistical mechanical approach for describing the
dynamic properties of a complex system, namely, a concentrated
suspension of yolk-shell particles. Here we have discussed the
simplest illustrative model representation, in which each shell
carries only a single yolk, and in which the yolk-shell and
shell-shell forces are model as purely repulsive, hard-body
interactions. This implied an additional simplification, namely,
the absence of yolk-yolk direct interactions. These
simplifications allowed us to reach a reasonable understanding of
the differences between a yolk-shell system and a suspension of
empty shell, and of the main trends observed upon the variation of
relevant parameters, such as the short-time dynamic asymmetry
parameter $\delta$ or the shell volume fraction $\phi$. The
message, however, is that the theoretical approach presented here
can be extended to consider other, more complex conditions, such
as including more than one yolks per shell and studying the
effects of yolk-shell and shell-shell interactions beyond the
purely repulsive, hard-core--like interactions considered here.

\section{Acknowledgments}
This work was supported by  the U.S. Department of Energy, Office
of Basic Energy Sciences, Materials Sciences and Engineering
Division. This Research at the SNS at Oak Ridge National
Laboratory was sponsored by the Scientific User Facilities
Division, Office of Basic Energy Sciences, U.S. Department of
Energy. This work was also supported by the Consejo Nacional de
Ciencia y Tecnolog\'ia (CONACYT, Mexico), through Grants No.
132540 and No. 182132.


\vskip3cm

\begin{thebibliography}{99}

\bibitem{lou} X.W. Lou, L. A. Archer, Z. Yang, Adv. Mater. 2008, \textbf{20}, 3987.

\bibitem{joo} S. H. Joo, J. Y. Park, C. K. Tsung, Y. Yamada, P. D. Yang, G. A.
Somorjai, Nat. Mater. 2009, \textbf{8}, 126.

\bibitem{yin} Y. D. Yin, R. M. Rioux, C. K. Erdonmez, S. Hughes, G. A.
Somorijai, A. P. Alivisatos, Science 2004, \textbf{304}, 711.

\bibitem{kam} K. Kamata, Y. Lu, Y. N. Xia, J. Am. Chem. Soc. 2003, \textbf{125},
2384.


\bibitem{liu} J. Liu, H. Xia, D. F. Xue, L. Lu, J. Am. Chem. Soc. 2009, \textbf{131},
12086.

\bibitem{li} H. X. Li, Z. F. Bian, J. Zhu, Y. N. Huo, H. X. Li, Y. F. Lu, J. Am.
Chem. Soc. 2007, \textbf{129}, 8406.





\bibitem{ermak} D. L. Ermak and J. A. McCammon. J. Chem. Phys. {\bf 69}, 1352
(1978).
\bibitem{dub}D. Dubbeldam, D. C. Ford, D. E. Ellis, and R. Q. Snurr,
Mol. Sim. {\bf 35}, 1084 (2009).



\bibitem{scgle0}  L. Yeomans-Reyna and M. Medina-Noyola, Phys. Rev. E {\bf
62}, 3382 (2000).

\bibitem{scgle1}  L. Yeomans-Reyna and M. Medina-Noyola, Phys. Rev. E {\bf
64}, 066114 (2001).

\bibitem{scgle2}  L. Yeomans-Reyna, H. Acu\~{n}a-Campa,
F. Guevara-Rodr\'{\i}guez, and M. Medina-Noyola, Phys. Rev. E {\bf
67}, 021108 (2003).


\bibitem{rmf}  P.E. Ram\'{\i}rez-Gonz\'alez {\it et al.}, Rev. Mex. F\'{\i}sica
\textbf{53}, 327  (2007).

\bibitem{todos1} L. Yeomans-Reyna, M. Chavez-Rojo, P.E. Ram\'{\i}rez-Gonz\'alez,R. Ju\'arez-Maldonado,M. Chavez-Paez and  M. Medina-Noyola  Phys. Rev. E \textbf{76},
041504 (2007).

\bibitem{todos2} R. Ju\'arez-Maldonado, M. Chavez-Rojo,P.E. Ram\'{\i}rez-Gonz\'alez,L. Yeomans-Reyna and  M. Medina-Noyola,  Phys. Rev. E {\bf 76}, 062502 (2007).

\bibitem{faraday}  M. Medina-Noyola, Faraday Discuss. Chem. Soc. {\bf 83},
21 (1987).
\bibitem{overdampedatomic} L. L\'opez-Flores, L. L. Yeomans-Reyna
and M. Medina-Noyola, J. Phys.: Condens. Matter \textbf{24},
375107 (2012).


\bibitem{percus}  J. K. Percus and G. J. Yevick, Phys. Rev. {\bf 110}, 1 (1957).
\bibitem{verlet} L. Verlet and J.-J. Weis,  Phys. Rev. A  {\bf  5} 939 (1972).

\bibitem{allen} Allen, M. P.; Tildesley, D. J. Computer Simulation of Liquids.
 Oxford University Press: Oxford, 1989.

 \bibitem{guev}  F. de J. Guevara-Rodr\'{\i}guez and M.Medina-Noyola,
 Phys. Rev. E {\bf   68}, 011405 (2003).

\bibitem{lety}L. Lopez-Flores, H. Ruiz-Estrada, M. Ch\'avez-P\'aez and M. Medina-Noyola.,Phys. Rev. E {\bf 88}, 042301 (2013)

\bibitem{paperlargo} L. E. S\'anchez-D\'iaz, E. C.
Cort\'es-Morales, X. Li, W.-R. Chen, and M.
Medina-Noyola, manuscript in preparation (2014).






\end{thebibliography}
\end{document}